\documentclass[twocolumn]{aastex631}  % linenumbers

\usepackage[capitalise]{cleveref}
\usepackage{comment}
\usepackage{graphicx}
\usepackage{appendix}
\usepackage{multirow}
\usepackage[]{footmisc}
\usepackage[caption=false]{subfig}

\shorttitle{Analysis of TeV-detected Gamma-Ray Bursts}

\shortauthors{L. Foffano, M. Tavani}

\begin{document}

\title{TeV Afterglows of Gamma-Ray Bursts:\\ Theoretical Analysis and Prospects for Future Observations}

\correspondingauthor{Luca Foffano}
\email{luca.foffano@inaf.it}

\author[0000-0002-0709-9707]{Luca Foffano}
\affiliation{INAF-IAPS Roma, via del Fosso del Cavaliere 100, I-00133 Roma, Italy}

\author[0000-0003-2893-1459]{Marco Tavani}
\affiliation{INAF-IAPS Roma, via del Fosso del Cavaliere 100, I-00133 Roma, Italy}

\received{December 20$^{\text{th}}$, 2024}
\accepted{August 5$^{\text{th}}$, 2025}

\begin{abstract}
Recent detections of  gamma-ray bursts (GRBs) at TeV energies opened new prospects for investigating radiative environments and particle acceleration mechanisms under extreme conditions. In this paper, we study the afterglows of these GRBs - namely GRB~180720B, GRB~190114C, GRB~190829A, GRB~201216C, and GRB~221009A - modeling their synchrotron and inverse Compton emission within the framework of an optimized relativistic fireball model. 
We constrain the model parameters and their temporal evolution by applying our theoretical model to the high-energy emission in the X-ray and GeV-TeV energy bands observed at intermediate and late times. 
Our results reveal interesting differences among the TeV-detected GRBs, potentially reflecting a variety of underlying physical processes that lead to different maximum energies $E_{\text{max}}= \, \gamma_{\text{max}}\, m_e \, c^2$ of the accelerated particles responsible for the GRB high-energy emission.
We indeed obtain different behaviors of the late TeV afterglows that ultimately depend on $\gamma_{\text{max}}$. We discuss how late afterglow observations - on timescales of hours and days - of X-ray and GeV-TeV emissions are crucial for providing diagnostics of the physical processes behind GRBs,  and we emphasize the theoretical expectations for future TeV observations.
\end{abstract}


%%%%%%%%%%%%%%%%%%%%%%%%%%%%%%%%%%%%%%%%%%%%%%%%%%%%%%%%%%%%%%%%%%%%%%%%%

\section{Introduction} 
\label{sec:intro}

\noindent
Gamma-ray bursts (GRBs) are the most energetic explosions in the Universe, and they represent a fascinating enigma in modern astrophysics. The origin of long-duration GRBs (lasting $>$2~s) is commonly associated with the core collapse of massive stars, while short-duration GRBs (lasting $<$2 s) are probably connected to the merging of compact objects, such as neutron stars.
These events generate collimated jets of plasma where particles are accelerated at ultra-relativistic velocities, requiring extreme particle acceleration mechanisms \citep[see e.g. a recent review by][]{2018IJMPD..2742003N}.

GRB emission typically begins with a short and intense prompt phase, followed by a long afterglow. 
The prompt phase can be very intense and it is usually characterized by emission in the X-ray/MeV energy range, mostly due to the dynamic activity of the central engine. The subsequent afterglow phase is mostly related to the propagation of the blast wave into the surrounding medium. 

The recent discovery of very-high energy (VHE, above 100 GeV) gamma rays from the afterglow of several GRBs has opened a new window in the theoretical and observational study of these transient events. 
The first ever TeV detection of a GRB dates back to 2018 with GRB~180720B, which was observed starting $10$~hours after the trigger event \citep{2019Natur.575..464A}. However, only in 2019 very early TeV radiation right at the onset of the afterglow phase - starting just one minute after the trigger time - was detected from a GRB with observations by the MAGIC telescopes of GRB~190114C  \citep{MAGIC_GRB190114C}. These breakthrough events marked a transition to a new era of investigation into the extremely energetic VHE radiation emitted by these phenomena, supporting an interpretation in terms of inverse Compton (IC) reprocessing of synchrotron photons \citep{2001ApJ...559..110Z} and highlighting the role of the Synchrotron Self-Compton (SSC) model to describe the emitted TeV photons \citep{sari_esin_2001}.

In the following years, the detection of further GRBs at VHE provided crucial information for the development of this field. It turned out that their diverse observational properties challenge the traditional GRB paradigm, showing flux evolutions and spectral behaviors that are not fully compatible with a global evolution expected by standard models. Understanding these differences and the underlying processes is crucial for exploring this new domain in GRB science and developing a comprehensive theoretical framework for these events.

In this work, we  investigate all GRBs currently detected at TeV energies, aiming at comparing their VHE emission and exploring their underlying properties. We  specifically discuss their intermediate and late-time high-energy flux evolution up to a few days after the trigger time, considering possible new features arising in their VHE emission. We adopt an optimized version of the standard relativistic fireball model \citep[e.g.,][]{Rees_1992_relativistic_fireball, 1997ApJ...476..232M}, in which - among other features - the particle acceleration mechanism is characterized by specific values of the maximum energy $\gamma_{\text{max}}$ of the accelerated particles (instead of the assumption $\gamma_{\text{max}} = \infty$ usually considered in the literature).\\

\noindent
This paper is organized as follows. In Section~\ref{sec:mwl_data}, we summarize the multi-wavelength (MWL) datasets adopted in this analysis for each GRB.
In Section~\ref{sec:modeling}, we briefly describe the theoretical model applied to the data, reporting in Section~\ref{sec:results} the resulting sets of parameters adopted for each event. Then, in Section~\ref{sec:discussion}, we discuss the results and compare the TeV gamma-ray light curves of our sample of GRBs, showing the possible rising phenomenology and future tests.

Throughout all the paper, all times are referred from the trigger times $T_0$ of each GRB indicated in Section~\ref{sec:mwl_data}. The only exception is GRB~221009A, which is shown from time $T' = T_0 + 226$~s, following \citet{LHAASO_2023} and coherently with our previous work \citep{foffano2024_GRB221009A}.
In our calculations, we assume cosmological parameters describing a flat Universe with $\Omega_{M}$ = 0.3, $\Omega_{\Lambda}$ =
0.7 and $H_{0}$ = 70~km s$^{-1}$~Mpc$^{-1}$.\\

\begin{table*}
    \caption{List of TeV gamma-ray detected GRB afterglow events. On the left part of the table, we indicate event name and redshift $z$. On the right, we report the parameter values adopted in the relativistic fireball modeling of these GRBs. In order we indicate the isotropic-equivalent initial energy $E_{\text{iso,0}}$, the initial bulk Lorentz factor $\Gamma_0$, the density profile index $s$ and normalization  n$_0$ or $A^*$ (in the case of homogeneous or wind-like medium, respectively), the electron distribution index $p$ and maximum energy $\gamma_{\text{max}}$, and the parameters $\varepsilon_e$ and $\varepsilon_B$. For GRB~221009A, we have adopted a set of time-evolving microphysical efficiencies, as reported in \citet{foffano2024_GRB221009A}. Considering the parameter correlations, we estimate the following model-driven uncertainties: 0.5 for the estimated $ \log_{10} E_{\text{iso,0}}$, 100 for $\Gamma_0$, 0.5 for n$_0$, 0.2 for $p$, 0.5 for $ \log_{10} \gamma_{\text{max}}$, 0.5 for $\varepsilon_e$ and 0.5 for $\varepsilon_B$.  }
    \renewcommand{\arraystretch}{1.5}
    \centering
    \hspace{-2cm}\begin{tabular}{lc|ccccccccc}
    \toprule
    Event & z & $ \log_{10} E_{\text{iso,0}}$ & $\Gamma_0$ & s & n$_0$ & $A^*$ & $p$ & $ \log_{10} \gamma_{\text{max}}$ & $\varepsilon_e$ & $- \log_{10} \varepsilon_B$ \\ 
    ~ &  & $- \log_{10}$[erg] & ~ & ~ & [cm$^{-3}$] & & ~ & ~ &  & ~ \\ \hline
    GRB~180720B & 0.65   & $54.8$ & 400 & 0 & 1.2 & - & 2.6 & $> 6.6$ & 0.28 & 6.6 \\ \hline 
    GRB~190114C & 0.42 &  $53.8$ & 500 & 0 & 0.2 & -& 2.8 & $9.0$ & 0.05 & 4.3 \\ \hline 
    GRB~190829A & 0.0785 &  $53.5$ & 500 & 0 & 1.0 & -& 2.4 & $8.8$ & 0.07 & 6.5 \\ 
    \hline
    GRB~201216C & 1.1 &    $53.8$ & 200 & 2 & -& 0.04 & 2.6 & $7.9$ & 0.045 & 2.4 \\ \hline 
    GRB~221009A & 0.151 &    $55.8$ & 480 & 0 & 0.8 & -& 2.5 & 7.3 & 0.017 $\mapsto$ 0.05   & $4.7\mapsto6.9$ \\ \hline 
    \toprule
    \end{tabular}
    \label{tab:list_grbs}
\end{table*}

%%%%%%%%%%%%%%%%%%%%%%%%%%%%%%%%%%%%%%%%%%%%%%%%%

\section{The sample of TeV-detected Gamma-ray bursts}
\label{sec:mwl_data}
\noindent
Our sample is composed by five GRBs detected at TeV gamma-ray energies up to mid 2025: GRB~180720B, GRB~190114C, GRB~190829A, GRB~201216C, and GRB~221009A. Their main observational properties are summarized in the next paragraphs, where we will describe the datasets adopted for the modeling of each GRB. Our study will focus on the available data at optical, X-ray, GeV, and TeV gamma-ray energies. We exclude radio data and very early optical and X-ray observations (as specified for each GRB), as these are likely influenced by a reverse shock or prompt contributions, which are intentionally not included in our model.\\

\subsection{GRB~180720B}
\label{sec:GRB180720B}
\noindent
GRB~180720B was first detected by \textit{Fermi}-GBM and \textit{Swift}-BAT on 2018 July 20$^{\text{th}}$ at $T_0=$~14:21:39.65 UT \citep{2018GCN.22981....1R, 2018GCN.22973....1S}, and classified as a long-duration GRB with redshift $z = 0.654$ \citep{2018GCN.22996....1V}. 

X-ray observations are available for the whole burst, including both the prompt and afterglow phases, extending up to a month after $T_0$.
The complex X-ray lightcurve, characterized by multiple peaks and flux variations, supports the presence of several phases of this GRB.  The X-ray flares up to $200$~s have been interpreted as possible interactions of a reverse shock expanding in a thin shell \citep{2019ApJ...885...29F}. 

In the high-energy (HE, energy between 100 MeV and 100 GeV) gamma-ray band, this GRB was also detected by the \textit{Fermi}-LAT telescope up to 700~s. 
No further HE emission was detected in the following observations.

GRB~180720B was detected by H.E.S.S. \citep{2019Natur.575..464A}, marking the first detection of VHE gamma~rays deep into a GRB afterglow phase. Its observations between 100–440 GeV began at $T_0$+10 h and lasted for two hours.

For the analysis of this GRB, we have considered the dataset reported in \citet{2019Natur.575..464A}, including the optical r~band, X-ray data provided by \textit{Swift}-XRT (0.3-10 keV), GeV data from \textit{Fermi}-LAT  (0.1-100 GeV), and finally H.E.S.S. data (100-440 GeV).\\

\subsection{GRB~190114C}
\label{sec:GRB190114C}
\noindent
On 2019 January 14$^{\text{th}}$ at $T_0=$~20:57:03.19 UT, the long GRB~190114C was detected by \textit{Swift}  and \textit{Fermi}  \citep{2019GCN.23707....1H, 2019GCN.23688....1G}, among other observatories, reporting  exceptional gamma-ray emission extending into the TeV range. Its redshift has been estimated as z = $0.4245 \pm 0.0005$ \citep{2019GCN.23708....1C}.
The light curves in X-rays and in low-energy gamma~rays for \textit{Swift}  and \textit{Fermi}  \citep{2020ApJ...890....9A}, and for AGILE and \textit{Konus}-WIND \citep{2020ApJ...904..133U} reported high rapid variability in the early prompt emission, which lasted for approximately 6~s. Additionally, a further reflaring episode in the X-ray band has been reported between 12 and 25 s, without apparent connected gamma-ray contribution.

This burst became the first confirmed GRB reaching the TeV regime, thanks to follow-up observations by MAGIC \citep{MAGIC_GRB190114C}, which observed it between $10^2$ and $10^3$~s.

In HE gamma~rays, this GRB was in the field of view of \textit{Fermi}-LAT for a long time between 150 and 8600~s, reporting an interesting coherent flux evolution in this energy band, similarly to the one reported in the X-ray band \citep{2019GCN.23709....1K, 2020ApJ...890....9A}. 

For this GRB, we have considered the data reported in \citet{MAGIC_GRB190114C}, including optical data, X-ray data provided by \textit{Swift}-XRT (1-10 keV), GeV data from \textit{Fermi}-LAT  (0.05-3 GeV), and finally MAGIC data (0.3-1 TeV).\\

\subsection{GRB~190829A}
\label{sec:GRB190829A}
\noindent
GRB~190829A was a remarkable long GRB detected on 2019 August 29$^{\text{th}}$ at $T_0=$~19:55:53 UT by \textit{Fermi}-GBM \citep{2019GCN.25551....1F} with redshift z = $0.0785 \pm 0.0005$ \citep{2021A&A...646A..50H}, 

The X-ray light curve reports a dynamical and active prompt emission up to thousands of seconds after $T_0$. After a final rebrightening at about 3000 seconds, the X-ray light curve landed on a typical power-law decay, indicating the onset of the afterglow phase. 

Most theoretical models of the first phases of the GRB indicate the presence of an evident emission produced by a reverse shock in addition to the forward shock emission up to $\sim 10^4$~s, after which the former contribution  becomes negligible \citep[e.g.,][]{2020MNRAS.496.3326R, 2022MNRAS.512.2337D, 2022ApJ...931L..19S}. 

{\textit{Fermi}-LAT did not detect any HE gamma-ray radiation from GRB~190829A, providing only flux upper limits at $\sim 10^4$~s \citep{GRB190829A_latupperlimits}.}

Conversely, this event was detected at VHE gamma~rays by H.E.S.S., which observed it on three consecutive nights, from 4.3 to 55.9 hours after $T_0$. 
The afterglow emission of GRB~190829A reported similar temporal properties across the X-ray and TeV bands, which were interpreted as extreme synchrotron radiation or a coexistence of synchrotron and inverse Compton processes \citep{2021Sci...372.1081H}.

For this GRB, we have considered the data reported in \citet{2021Sci...372.1081H}, including X-ray data provided by \textit{Swift}-XRT (0.3-10 keV), GeV upper limits from \textit{Fermi}-LAT  (0.1-100 GeV), and finally H.E.S.S. data (0.2-4 TeV).\\

\subsection{GRB~201216C}
\label{sec:GRB201216C}
\noindent
GRB~201216C was detected in X-rays on 2020 December 16$^{\text{th}}$ at $T_0=$~23:07:31 UT by \textit{Swift}-BAT \citep{swift_GRB201216C}, and later also by \textit{Fermi}-GBM, ASTROSAT, and Konus-Wind. A set of MWL observations revealed extended high-energy emission persisting for thousands of seconds, reporting a complex lightcurve.

This event has pushed the limits of the maximum distance of a TeV-detected GRB, reporting a redshift of $z = 1.1$ \citep{2020GCN.29077....1V}.
The TeV detection of such high-redshift event has important implications on the studies on EBL absorption. 

The MAGIC telescopes observed the GRB after about 100 seconds, reporting a clear detection after 20 minutes of observations \citep{MAGIC_GRB201216C}. 
Conversely, \textit{Fermi}-LAT data observed this event between 3500-5500~s, but did not report any detection, providing just a flux upper-limit \citep{fermi_GRB201216C}. No simultaneous data were registered in the gamma-ray and X-ray band, with \textit{Swift}  that observed the event only after 2967~s and up to 4325~s \citep{swift_GRB201216C}.
The GRB was  also observed by HAWC between 100 and 3600~s, reporting no signal detection \citep{2020GCN.29086....1A}.

For this GRB, we have considered the data reported in \citet{2020GCN.29077....1V}, including optical r'~band, X-ray data provided by \textit{Swift}-XRT (0.3-10 keV), GeV data from \textit{Fermi}-LAT  (0.1-1 GeV), and finally MAGIC data (70-200 GeV).

\subsection{GRB~221009A}
\label{sec:GRB221009A}
\noindent
GRB~221009A is recorded as the brightest transient event of all times. It was detected on 2022 October 9$^{\text{th}}$ at $T_0=$~13:16:59.99~UT by \textit{Swift}  \citep[J1913.1+1946,][]{2022ATel15651....1N} and \textit{Fermi}  \citep{2022GCN.32636....1V, fermi_gbm_grb221009a}. It is attributed to the core collapse of a massive star \citep{Srinivasaragavan2023, supernova_grb221009a}, located at a redshift of $z = 0.15095 \pm 0.00005$ \citep{2023arXiv230207891M, 2022GCN.32648....1D}. Its prompt emission phase produced a peak photon flux so high that most space-based detectors experienced saturation, revealing an unprecedented intense brightness in X-rays and gamma~rays. 

In addition, GRB~221009A reported an extended energetic emission covering all energy bands and reaching the highest TeV energies ever observed from any GRBs. The detection of TeV photons by the LHAASO observatory \citep{LHAASO_2023, LHAASO_data_above_10tev} opened a new perspective for the study of particle acceleration in GRBs.

Interestingly, new results from the first Large-Sized Telescope (LST-1) of the Cherenkov
Telescope Array Observatory \citep[CTAO,][]{Abe:2023Vc} have reported a hint of detection of GRB~221009A at a $4.1\sigma$ significance level, occurring  approximately after $\sim$$10^5$~s  after the trigger \citep{Abe_2025, LST_GRB221009A_hepro, LST_GRB221009A, LST_GRB221009A_memsait}. This result is particularly interesting since it may rule out standard scenarios of GRB afterglow evolution, and align with a possible jet break involving the TeV gamma-ray emission \citep[e.g.,][]{2024JHEAp..41...42Z, oconnor_structured_jet, LHAASO_2023} or represent the effect of the spectral signature of the maximum energy of the particle distribution \citep{foffano2024_GRB221009A}.

For this GRB, we have considered optical data obtained by \citet{Fulton_optical_lightcurve}, X-ray data provided by XRT (0.3-10 keV) and BAT (14-195 keV) instruments on board the \textit{Swift} telescope \citep{Swift_williams_2023}, GeV gamma-ray data from AGILE GRID \citep[0.05 - 3 GeV,][]{Tavani_2023} \textit{Fermi}-LAT \citep[0.1 - 100 GeV,][]{Axelsson_2025}, an upper-limit at TeV energies by HAWC \citep[0.3-5 TeV,][]{2022GCN.32683....1A}, and finally 
 the TeV gamma-ray LHAASO data \citep[0.3 - 7 TeV,][]{LHAASO_2023, LHAASO_data_above_10tev}.

\subsection{Marginal TeV detections of GRB~201015A and GRB~160821B}
\noindent
In addition to the five GRBs reporting TeV gamma-ray detection mentioned above, other two GRBs have been claimed as hints of detection in this energy band, namely GRB~201015A and GRB~160821B.

In the case of GRB~201015A, it was first discovered by \textit{Swift}-BAT on 2020 October 15$^{\text{th}}$ at $T_0=$~22:50:13~UT \citep{swift_GRB201015A}. Later on, a  hint of detection of $>3\sigma$ was announced by the MAGIC Collaboration with a notification via General Coordinates Network \citep[GCN,][]{2020GCN.28659....1B}. Since there are not available datasets for this event, we exclude it from our analysis.

Conversely, GRB~160821B was studied in detailed by the MAGIC Collaboration, reporting just a hint of detection of $>3\sigma$ \citep{2021ApJ...908...90A}. It is identified as a short GRB and first identified by \textit{Swift}  on 2016 August 21$^{\text{st}}$ at 22:29:13~UT, reporting a redshift of $z = 0.162$ \citep{swift_GRB160821B, redshift_GRB160821B}.
Since no clear detection in the VHE band has been provided also for this event, we exclude it from our dataset. \\

\begin{figure*}
    \centering
    \subfloat[Light curve.]{\includegraphics[width=0.7\textwidth]{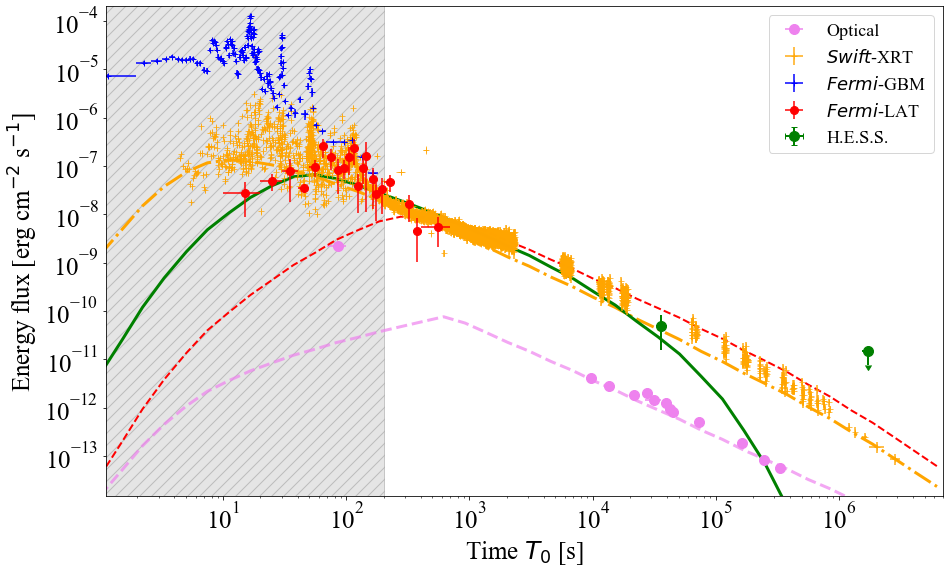}\label{fig:lightcurve_GRB180720B}}\\
    %\hspace{0.5cm}
    \subfloat[Spectrum.]{\includegraphics[width=0.7\textwidth]{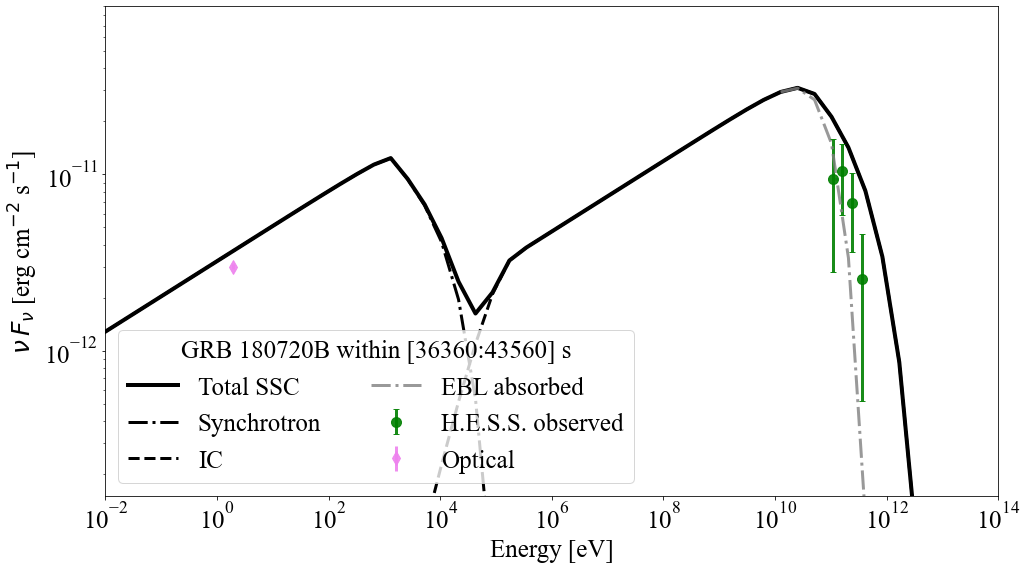}\label{fig:sed_GRB180720B}}
    \caption{ \textit{(Left panel)} Observed and calculated flux temporal evolution of GRB~180720B within the global modeling described in Section~\ref{sec:modeling}. For all plots, we have adopted the following color code: violet for optical data, orange for X-ray data, red for gamma-ray data, and green for TeV data.  The X-ray data are well matched by our model throughout the whole evolution of the GRB afterglow; optical and HE gamma-ray data are also well reproduced. The shaded areas indicate early times during which prompt-like dynamical activity influences the emission. This phase is not intended to be described by the model; however, we notice that our model evolution at early times matches well with that  data. A late-time steepening in the X-ray and gamma-ray bands is due to the crossing of the critical cooling frequency $\nu_c$ and the role of the maximum electron energy, $\gamma_{\text{max}}$, which is crucial for matching the late-time TeV gamma-ray data. \textit{(Right panel)}
    The calculated broad-band spectrum at the time of the TeV detection within $\Delta t = 36300-43560$~s after trigger. We show the H.E.S.S. spectral data and optical data. }
    \label{fig:lightcurve_and_sed_GRB180720B}
\end{figure*}

\begin{figure*}
    \centering
\subfloat[Light curve.]{\includegraphics[width=0.7\textwidth]{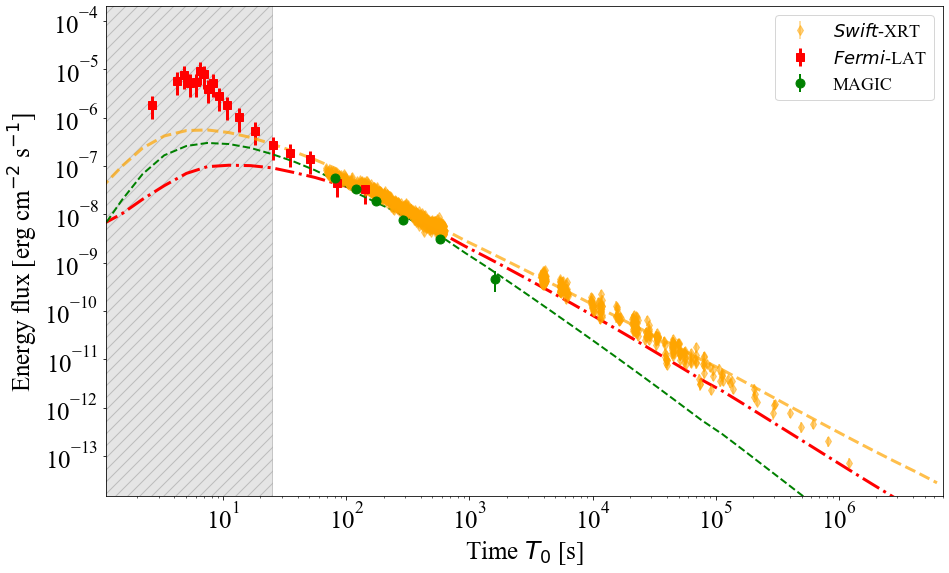}
    \label{fig:lightcurve_GRB190114C}}\\
\hspace{0.5cm}
    \subfloat[Spectrum.]{\includegraphics[width=0.7\textwidth]{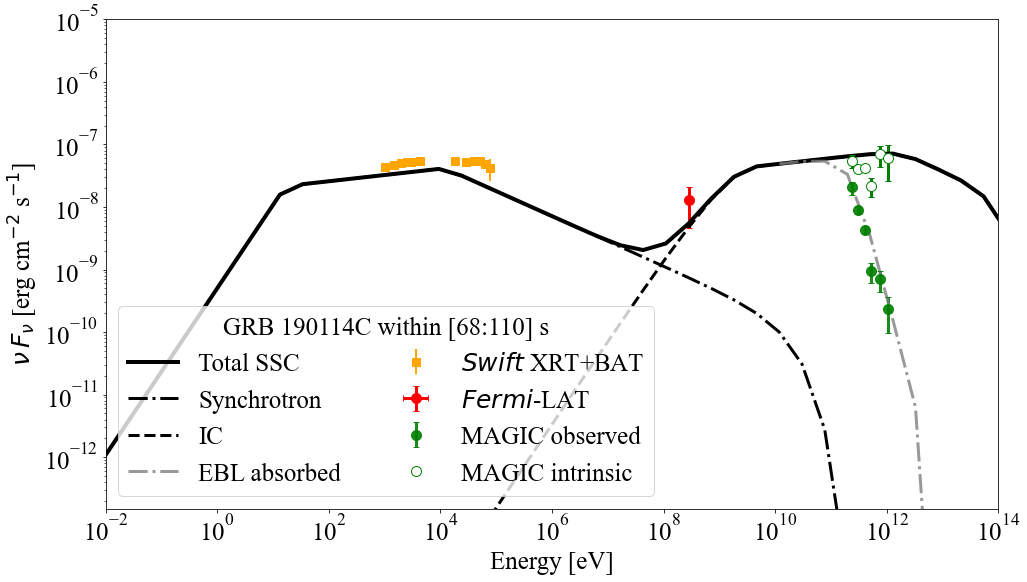}\label{fig:sed_GRB190114C}}
    \caption{Similarly to Figure~\ref{fig:lightcurve_and_sed_GRB180720B}, for GRB~190114C. \textit{(Left panel)} Observed and calculated  light curves: our model accurately matches the X-ray and TeV gamma-ray bands throughout the GRB's evolution, while HE gamma-ray data are well reproduced after the prompts phase. \textit{(Right panel)} 
    The calculated broad-band spectrum at the onset of the TeV detection in the time interval $\Delta t = 68 - 110$~s after trigger.
    In the spectrum, we use the dataset reported in \citet{MAGIC_GRB190114C}, and include data of \textit{Swift}-XRT and BAT, and \textit{Fermi}-LAT.}
\end{figure*}

\begin{figure*}
    \centering
\subfloat[Light curve.]{\includegraphics[width=0.7\textwidth]{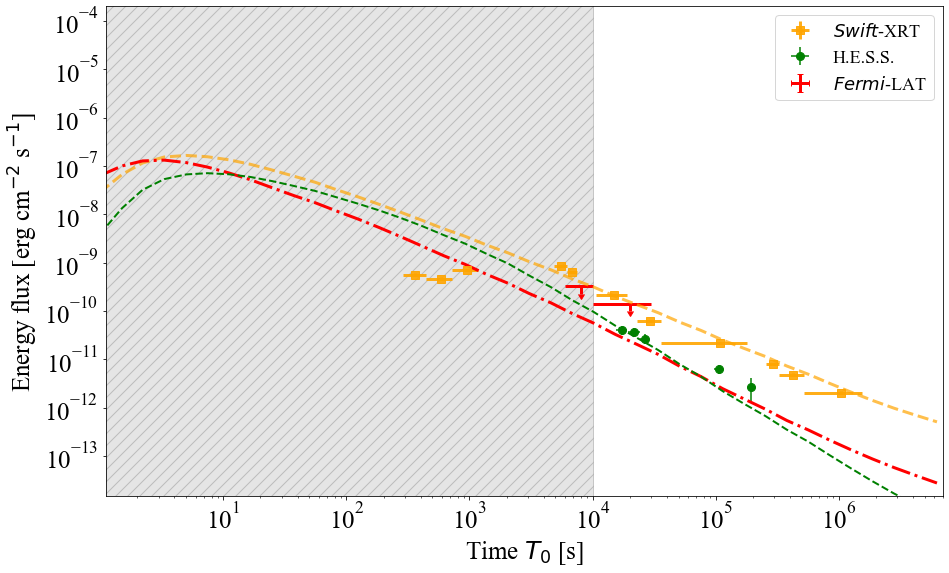}
    \label{fig:lightcurve_GRB190829A}}\\
%\hspace{0.5cm}
    \subfloat[Spectrum.]{\includegraphics[width=0.7\textwidth]{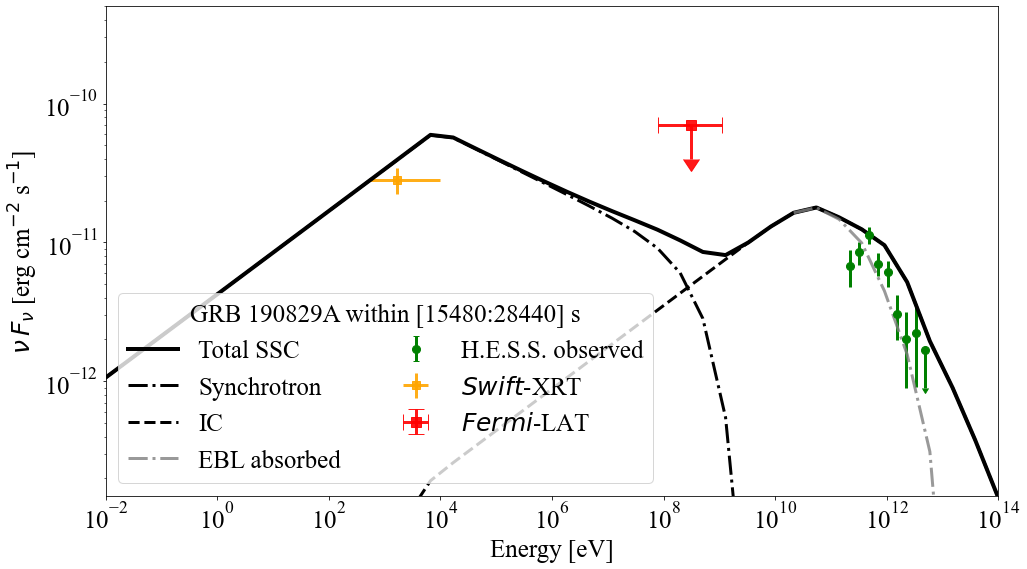}\label{fig:sed_GRB190829A}}
    \caption{Similarly to Figure~\ref{fig:lightcurve_and_sed_GRB180720B}, for GRB~190829A. \textit{(Left panel)} Light curves:  the X-ray and TeV gamma-ray bands are well matched by our model along the whole time evolution of the GRB. The former are not expected to be reproduced during the very early phase of  the prompt activity. The HE gamma-ray upper limits are consistent with our modeling. \textit{(Right panel)} 
     The calculated broad-band spectrum at the time  of the TeV detection in the time interval $\Delta t = 15480 - 28440$~s after trigger.
    In the spectrum, we use the dataset reported in \citet{2021Sci...372.1081H}, and include data of \textit{Swift}-XRT and BAT, and \textit{Fermi}-LAT. }
\end{figure*}

\begin{figure*}
    \centering
\subfloat[Light curve.]{\includegraphics[width=0.7\textwidth]{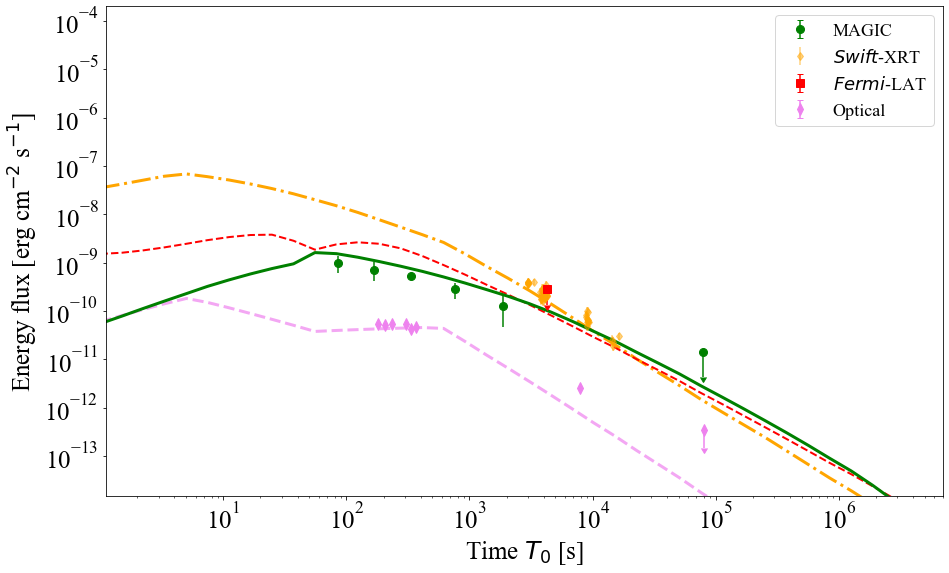}
    \label{fig:lightcurve_GRB201216C}}\\
%\hspace{0.5cm}
    \subfloat[Spectrum.]{\includegraphics[width=0.7\textwidth]{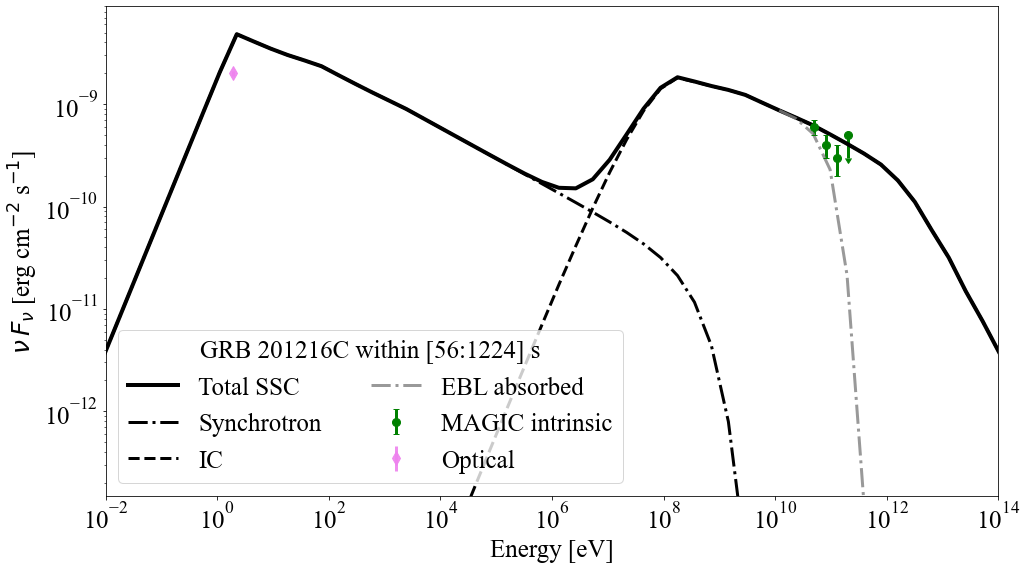}\label{fig:sed_GRB201216C}}
    \caption{Similarly to Figure~\ref{fig:lightcurve_and_sed_GRB180720B}, for GRB~201216C. \textit{(Left panel)} Light curves:  The X-ray and TeV gamma-ray bands are well matched by our model as whole time evolution of the GRB, together with the HE gamma-ray upper limit. Our model reproduces well the optical flux at early times,  and underestimates it at later times. \textit{(Right panel)} Calculated broad-band spectrum during the time of the TeV detection within $\Delta t = 56 - 1224$~s after trigger.
    We show the EBL-deabsorbed intrinsic MAGIC data, and the average optical spectral point.}
\end{figure*}

\begin{figure*}
    \centering
\subfloat[Light curve.]{\includegraphics[width=0.7\textwidth]{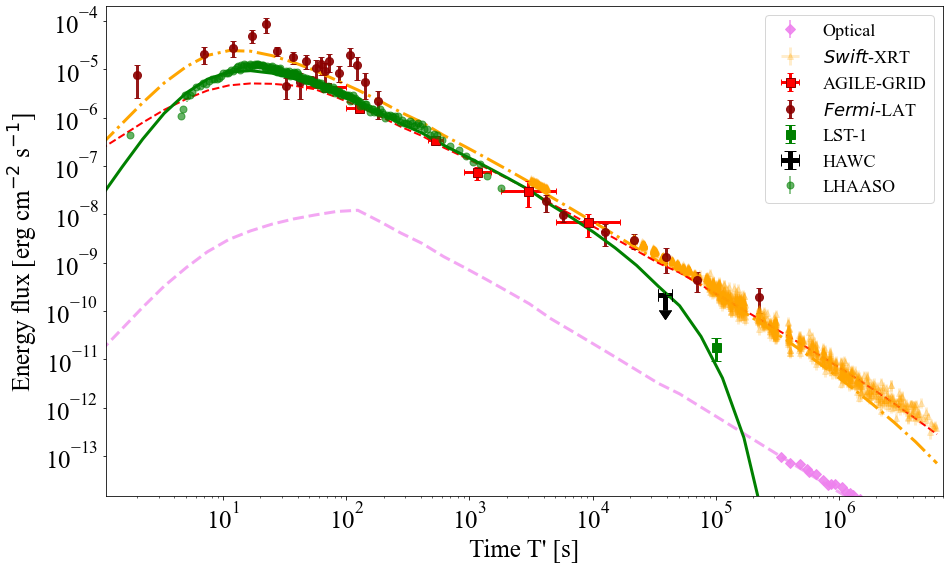}
    \label{fig:lightcurve_GRB221009A}}
\hspace{0.5cm}
    \subfloat[Spectrum.]{\includegraphics[width=0.7\textwidth]{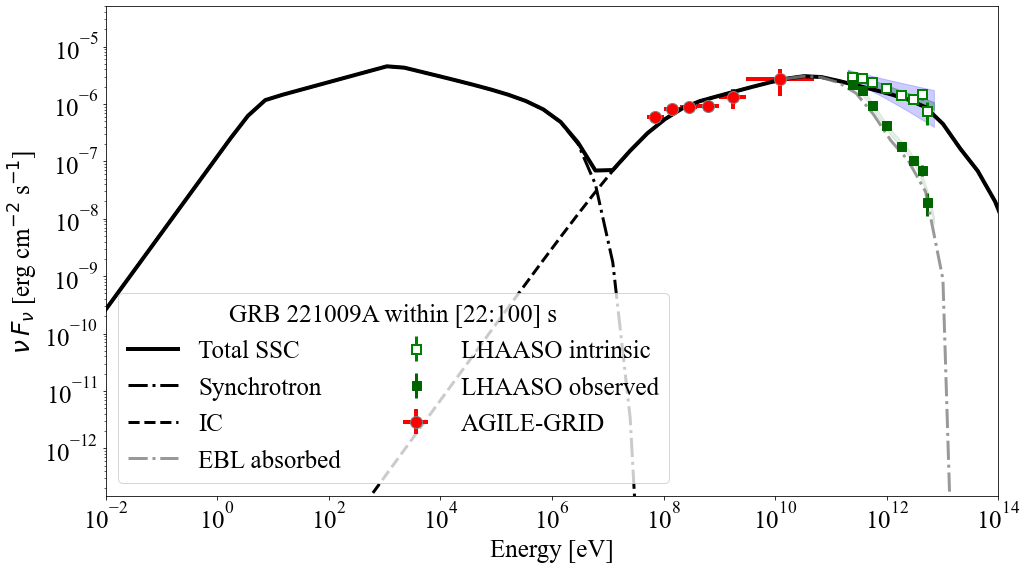}\label{fig:sed_GRB221009A}}
    \caption{Similarly to Figure~\ref{fig:lightcurve_and_sed_GRB180720B}, for GRB~221009A. \textit{(Left panel)} Light curves:
    our model reproduces well the time evolution of the flux light curves in different energy bands for the afterglow phase starting near $T' = 100$~s.
    Introducing a finite $\gamma_{\text{max}}$ is crucial to reproduce the TeV gamma-ray (and the X-ray) steepening at late times. \textit{(Right panel)} 
     Calculated broad-band spectrum at the time of the early GeV and TeV detections within $\Delta T' = 22 - 100$~s after trigger, for $T' = T_0 + 226$~s. The calculated spectrum shows the SED reported in \citet{foffano2024_GRB221009A} with simultaneous data of AGILE-GRID and LHAASO \citep{LHAASO_2023}.}
    \label{fig:lightcurve_and_sed_GRB221009A}
\end{figure*}

%%%%%%%%%%%%%%%%%%%%%%%%%%%%%%%%%%%%%%%%%%%%%%%%%%%
\section{Theoretical modeling}
\label{sec:modeling}
\noindent
In this paper, we apply a modified and optimized version of the \textit{relativistic fireball model}, aiming at modeling and comparing the five TeV-detected GRBs reported above. 

In the standard relativistic fireball model \citep[e.g.,][]{Rees_1992_relativistic_fireball, 1997ApJ...476..232M, 1998ApJ...499..301M, Piran_review_GRB_1998, Chiang_dermer_1999, Panaitescu_2000A}, the afterglow emission is produced by the interaction between a relativistic blast-wave - driven by jets produced during the core-collapse of massive stars - and the external environment. 
Forward shocks are produced and move outward in the external medium, while a reverse shock propagates inward into the shell. 
The expansion of the forward shock is assumed to be spherical, with an initial bulk Lorentz factor $\Gamma_0$ and an adiabatic hydrodynamic evolution $\Gamma(t)$ \citep{Blandford_1976, sari_hydrodynamics}. In the observer's frame, time and radial distance $r$ are connected by $r = 4 \Gamma^2 c t$ \citep{sari_1998}.

The shock front accelerates particles of the surrounding medium (ISM), which can be described by a density profile $n(r) \propto \, r^{-s}$ in the observer's frame. When the  blast wave is propagating into a homogeneous environment, we adopt $s = 0$ and a normalization constant $n_0$. When assuming a massive star wind-like medium, we use $s = 2$  and a normalization constant typically described with $A = 3 \times 10^{35}\, A^*$ cm$^{-1}$.

The energy of the shock $U$ is quickly split into magnetic field energy $U_B = \varepsilon_B U$ and random kinetic energy of electrons $U_e = \varepsilon_e U$ via the parameters $\varepsilon_B$ and $\varepsilon_e$, respectively.
The kinetic energy of the blast wave $E_{k}$ is carried mostly by the protons.
Particles of the external medium are accelerated by the shock, establishing a power-law energy distribution $dN (\gamma) / d\gamma \propto \gamma^{-p}$, with $p$ the power-law index, and $\gamma_{\text{min}}$ and $\gamma_{\text{max}}$ the two extreme energies of the particle distribution. 

{While in the standard relativistic fireball model $\gamma_{\text{max}} = \infty$ and the quantities $\varepsilon_B$ and $\varepsilon_e$ are constant, in our model,  depending on the specific GRB event:
\begin{itemize}
    \item we introduce a finite $\gamma_{\text{max}}$ quantity, which is ultimately a consequence of the efficiency of the particle acceleration mechanism;
    \item we consider $\varepsilon_B$ and $\varepsilon_e$ as possibly subject to time variation, following a defined temporal evolution.
\end{itemize}
Introducing these further effects with respect to the standard relativistic fireball model allows us to apply a more realistic approach that takes into consideration a physical evolution of the system.}

Two quantities define the physical regime of the system, namely $\gamma_{\text{min}}$ and the cooling Lorentz factor $\gamma_c = \frac{6 \pi m_e c}{\sigma_T \Gamma B^2 t}$. When $\gamma_{\text{min}} > \gamma_c$, particles are in a \textit{fast-cooling} regime, and loose efficiently their energy through synchrotron cooling within a dynamical time. On the other hand, when $\gamma_{\text{min}} < \gamma_c$, particles are in a \textit{slow-cooling} regime, and only particles with $\gamma > \gamma_c$~suffer efficient cooling. In our models, all events are described by this regime in the considered times.

The accelerated particles are then radiating their energy by synchrotron and inverse Compton emission through the SSC process \citep[e.g.,][]{sari_1998, sari_esin_2001}.

In our model, we take into consideration the possible internal $\gamma\gamma$ absorption, cosmological effects, and corrections for Klein-Nishina scattering \citep{Nakar2009}. Additionally, the absorption of gamma rays due to interaction with EBL photons is considered by adopting the model by \citet{ebl_dominguez_2011}.
The cooling effect due to the inverse Compton process that shortens the electrons' cooling time is considered in the model. The previously-defined cooling Lorentz factor is modified as $\gamma_c = \gamma_{\text{c,syn}} / (1+Y)$, where $Y$ is the Compton parameter computed following \citet{sari_esin_2001}.

It is important to note that, due to the inter-dependencies between the parameters of the adopted model, there is an unavoidable correlation in the chosen parameter space, which we take into account in the discussion.\\

%%%%%%%%%%%%%%%%%%%%%%%%%%%%%%%%%%%%%%%%%%%%%%%%%%%%%%%%%%
\section{Results}
\label{sec:results}\label{sec:SEDs}

\noindent
We applied the relativistic fireball model to the datasets of five TeV-detected GRBs presented in Section~\ref{sec:mwl_data}, with the primary goal of providing a robust description of the TeV gamma-ray data through forward-shock emission, as well as the best test of the model against the data in the optical, X-ray, and HE gamma-ray bands. A summary of the results and an estimation of the  best model parameters is reported in \Cref{tab:list_grbs}. The theoretical lightcurves, properly integrated over the same energy bands of the corresponding flux datasets available for these GRBs\footnote{{For the optical r band, we have integrated a typical energy range between 1.7 and 2.1 eV.}}, are reported in \Cref{fig:lightcurve_and_sed_GRB180720B}-\ref{fig:lightcurve_and_sed_GRB221009A}. The calculated SEDs taken at relevant times, which confirm an adequate match of the model to the corresponding available spectral datasets, are reported in the right panels of the same figures.

\begin{figure*}
    \centering
    \includegraphics[width=0.7\textwidth]{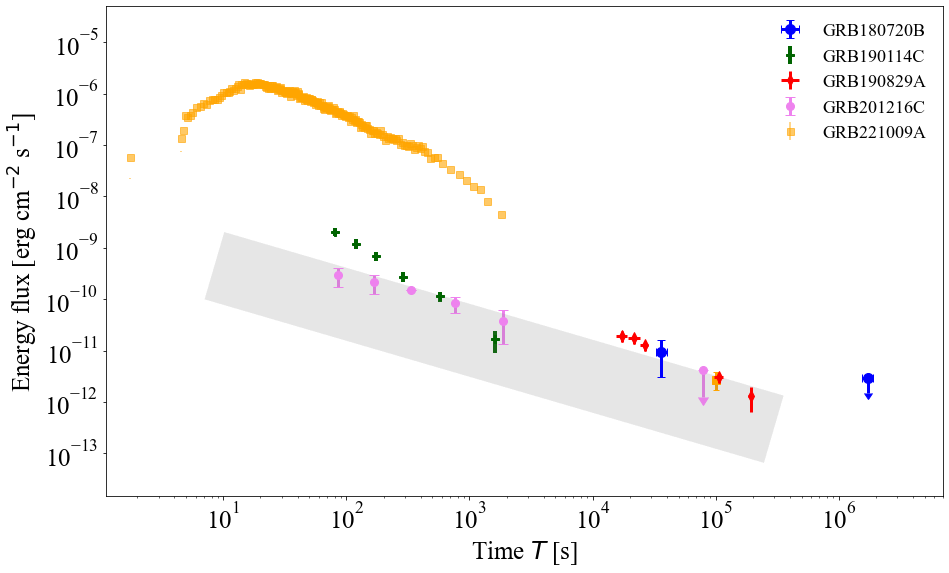}
    \caption{TeV gamma-ray datasets reported originally in \Cref{fig:lightcurve_and_sed_GRB180720B} and absorbed for the EBL with the model by \citet{ebl_dominguez_2011} to show the observed flux of the events. In this plot, observational data are reported from the trigger time $T \equiv T_0$ for all GRBs, except for GRB~221009A for which $T \equiv T' = T_0 + 226$~s. The shaded gray band indicates the interesting flux range which will be observable by the future generation of TeV Cherenkov telescopes.   }
    \label{fig:lightcurve_comparisons_onlydata_eblabs}
\end{figure*}

{We begin our discussion  by highlighting the exceptional nature of GRB~221009A. As illustrated in \Cref{fig:lightcurve_comparisons_onlydata_eblabs}, which shows the observed (EBL-absorbed) TeV light curves  for each GRB  presented in \Cref{fig:lightcurve_and_sed_GRB180720B}, this event was by far the brightest one ever detected  in the TeV range}, exceeding the observed  emission from all other GRBs by several orders of magnitude.
For this reason, it requested a special discussion that we addressed in \citet{foffano2024_GRB221009A}. In that work, we interpreted the multi-wavelength data of GRB~221009A within the relativistic fireball scenario modified by time evolution of the two microphysical parameters $\varepsilon_e$ and $\varepsilon_B$.  This modification was necessary and justified by the exceptional  spectral and evolution properties of the available data. Conversely, for the analysis of the other GRBs in this sample, we have adopted a standard relativistic model with constant microphysical parameters. 

After the publication of our previous work, new data have been reported, allowing us to update the light curves presented in \Cref{fig:lightcurve_and_sed_GRB221009A}. Specifically, the \textit{Fermi}-LAT instrument took data of GRB~221009A during the late afterglow between $10^3$~s and  $10^5$~s \citep{Axelsson_2025}. Such \textit{Fermi}-LAT points confirm the \textit{AGILE}-GRID datasets and the straight flux evolution of the afterglow at GeV energies up to very late times \citep{Tavani_2023}, in agreement with what expected by our model. Additionally, the smooth and continuous evolution of the GeV light curves during the afterglow challenges scenarios invoking a early-jet break at some hundreds seconds after trigger, requiring more complex interpretations.

{Moreover, further new datasets provide relevant constraints on $\gamma_{\text{max}}$}, which we identify in the late LHAASO data, the LST-1 hint of detection, and the HAWC flux upper limit.
In our refined model of GRB~221009A, which mostly follows what extensively discussed in \citet{foffano2024_GRB221009A}, we interpreted the likely TeV gamma-ray flux steepening - described by the three mentioned late-time datasets - as the effect of a spectral signature of the maximum energy of the particle distribution.
Our model remains largely consistent with all these observational results, and intriguingly suggests that HAWC was close to detecting a potential signal from this event.

Also in the case of GRB~180720B, the available datasets provide moderate constraints on $\gamma_{\text{max}}$. 
As mentioned in Section 2.1, the dynamical activity of GRB~180720B up to $200$~s has been attributed to reverse shock interactions, and therefore it has been excluded by our study. Just after this phase, our model shows a good match with the X-ray flux evolution. The HE gamma-ray emission is not reproduced before $200$~s, but the overall time evolution is well aligned with the latest data available in this band just exiting the dynamical prompt-like phase.
A chromatic break at $>10^5$~s is identified in X-rays (and HE gamma~rays, even without any available data to compare to) and not in optical, possibly indicating the transition of the cooling break in this energy band \citep{2022ApJ...931L..19S}. Indeed, in our model, the critical frequency also passes through the X-ray band at those times, as shown in Appendix~\ref{sec:appendix_nu_c}. However, its contribution to the integral flux in this energy range is not sufficient to explain the steepening observed after $10^5$ seconds, indicating that an additional effect must be considered. To address this  point, we adopted a $\gamma_{\text{max}}$ as low as $4\cdot 10^6$, which contributes to the X-ray bending. This feature also affects to the TeV gamma-ray band. 
The global modeling in X-rays, optical, and gamma~rays, forces the TeV gamma-ray band to be at higher fluxes during most of the decay, and to bend at $10^4$~s to match the late-time H.E.S.S. data point.  However, the correlation between model parameters prevents us to provide stronger constraints on the value of $\gamma_{\text{max}}$  as the origin of such steepening of the TeV flux, and in \Cref{tab:list_grbs} we indicate it 
as a lower limit. Lower values of $\gamma_{\text{max}}$ would lead to an earlier TeV bending, going against the data, but higher values of $\gamma_{\text{max}}$ may also be compatible in the case of alternative parameter configurations within the uncertainties. Our results are also in substantial agreement with \citet{Wang_2019} and \citet{2019ApJ...885...29F}, taking into consideration the expected degeneracy in the model parameter space.

\begin{figure*}
    \centering
    \includegraphics[width=0.7\textwidth]{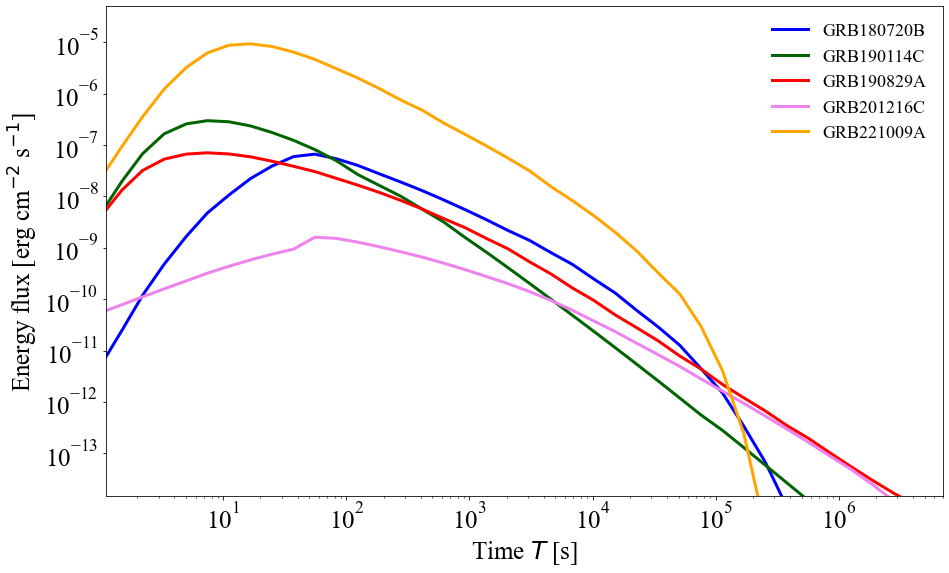}
    \caption{TeV gamma-ray calculated light curves for the considered GRBs (EBL de-absorbed with the model by \citealp{ebl_dominguez_2011}). The chosen TeV gamma-ray datasets are described in Section~\ref{sec:mwl_data}. Details of the parameter values for the model are reported in \Cref{tab:list_grbs}. Also in this plot, for all GRBs $T \equiv T_0$, except for GRB~221009A for which $T \equiv T' = T_0 + 226$~s.   }
    \label{fig:lightcurve_comparisons_new_nodata}
\end{figure*}

In the case of GRB~190114C, despite differences in modeling, the adopted parameters are in line with those reported in \citet{MAGIC_GRB190114C} and \citet{Derishev_2019}. Our model correctly reproduces the X-ray and the VHE gamma-ray emissions, both in the light curves and in the spectra. In the lightcurve, the model underestimates the HE gamma-ray emission during the first tens of seconds up to the X-ray rebrightening phase at 12-25~s - which is not intended to be modeled - but the  calculated flux evolution matches the data after about 50~s.  In the spectra, our model is below the hard X-ray data, which would probably be better interpreted by a smoother particle energy distribution, for example the one by \citet{MAGIC_GRB190114C}. On the other hand, our model succeeds in reproducing the rather flat spectral index in the considered GeV band, which we interpret as a consequence of the transition between synchrotron and IC emissions.

In the modeling of GRB~190829A, the resulting model parameters are in agreement with the current literature, and the assumed electron index $p\sim2.4$ is coherent with the softer value reported by \citet{2022MNRAS.512.2337D} with respect to the hard value $p\sim2.1$ reported by H.E.S.S.. As mentioned in Section 2.3, we do not expect to match the data before $\sim 10^4$~s, as they are probably contributed by substantial reverse shock emission.

Uniquely for the analysis of GRB~201216C, a wind-like external medium is better supported by the data due to the steeper time evolution of the flux data. For this event, which reports the highest redshift ever measured for a GRB detected at TeV energies, the best model has a quite low initial bulk Lorentz factor $\Gamma_0 = 200$ and a density  parameter $A^* = 0.04$. These results corroborate the findings of  \citet{MAGIC_GRB201216C}, although we adopt a more typical soft electron distribution with index $p = 2.6$. This value is also consistent with the physical parameters reported in \citet{2022ApJ...931..150H}, even if they adopt a different scenario. In our scenario, the optical flux evolution is reproduced by our model, even though reporting a moderate flux underestimation at $10^4$~s.
For the other GRBs, a homogeneous external medium successfully reproduces the data, resulting in parameter values in agreement with the current literature of GRBs.

%%%%%%%%%%%%%%%%%%%%%%%%%%%%%%%%%%%%%%%%%%%%
\section{Discussion}
\label{sec:discussion}

\noindent
In \Cref{fig:lightcurve_comparisons_new_nodata} we show 
our calculated de-absorbed TeV light curves for the five GRBs.
Each light curve is described by  different time dependencies reflecting the specific characteristics of the external medium and the best-choice model parameters. One interesting outcome of our analysis is the presence of a TeV steepening at times $10^4 < t < 10^6$~s in some events. 
While a late-time curvature of the TeV  light curve of GRB~221009A is evident both in the data and in the model, in GRB~180720B the only TeV data point detected by H.E.S.S. may provide a hint of curvature based on the specific modeling we have applied.
Conversely, this effect is clearly not supported by the data in GRB~190829A - which are not compatible with a late-time flux cutoff - and is not detectable even in the modeling of GRB~201216C and GRB~190114C extrapolating the available datasets at medium times. 
This TeV curvature may be due to different effects, such as a jet break \citep{Sari_jet_breaks}, a steepening due to the break of the high-energy components of a structured jet \citep[e.g.,][]{oconnor_structured_jet}, or  - as applied in this work - to the effect of a spectral curvature due to the presence of a maximum energy for the particle distribution that contributes to the integral flux in the energy band considered at the specific time \citep{foffano2024_GRB221009A}. 

{ The value of $\gamma_{\text{max}}$ of the radiating electrons ultimately depends on the details of the particle acceleration mechanism in the afterglow phase. A thorough discussion of this point for our sample of GRBs is postponed to future investigations.  However, we provide here some preliminary considerations to guide the interpretation. Of the two main acceleration mechanisms considered for GRB afterglows - collisionless shock acceleration \citep[e.g.,][]{Bell1978, Blandford1987} and magnetic field reconnection \citep[e.g.][]{Werner2017} - we consider here the case of  acceleration in electron-ion shocks following the analysis of \citet{Sironi2013}.
In this scenario, particles are accelerated by a diffusion process induced by Weibel-driven instabilities of scale $\lambda$ whose magnetic field $B$ ultimately sets the parameter $\epsilon_B = B^2/8 \, \pi \, \Gamma \, n \, m_i \, c^2$.
Considering that the time to reach the particle energy $\epsilon$ in relativistic shocks is $ t \sim D(\epsilon)/c^2$, with $D$ the diffusion coefficient $D \sim c \, \lambda \, (r_L/\lambda)^2$ and $r_L = \epsilon/e\, B$ the Larmor radius,
the maximum particle Lorentz factor can be expected to satisfy $\gamma_{\text{max}} \, m_i \, c^2 \sim e \, B \, \lambda \, (c\, t/\lambda)^2$. We then obtain the interesting relation $\gamma_{\text{max}} / \Gamma = (2 \epsilon_B \, \lambda \, \omega_{pi} \, t)^{1/2}$ (with $\omega_{pi}$ the ion plasma frequency). This relation elucidates the different factors influencing $\gamma_{\text{max}}$, i.e., 
the size and strength of the magnetic instabilities providing the necessary diffusion and the time of acceleration.}

{ We have verified that the $\gamma_{\text{max}}$ values obtained in our analysis are consistent with the expectations of collisionless shocks once the role of synchrotron cooling is taken into account. 
Our simplified considerations indicate the main physical quantities that ultimately determine the value of $\gamma_{\text{max}}$. Obtaining definite information on $\gamma_{\text{max}}$ goes to the heart of the theoretical modeling of GRB afterglows.}

The appearance of the TeV steepening only in particularly luminous GRBs such as GRB~221009A and potentially also on GRB~180720B may be related with the extreme energetics of these events  and the details of the magnetic structure of the particle acceleration front.
This feature suggests that only late-time observations of GRBs up to times of a few days may elucidate this point, helping to understand if the TeV light curve presents a steepening at late times, with a modification of the time dependence of the most energetic energy band.

\begin{figure}
    \centering
    \subfloat[]{\includegraphics[width=\columnwidth]{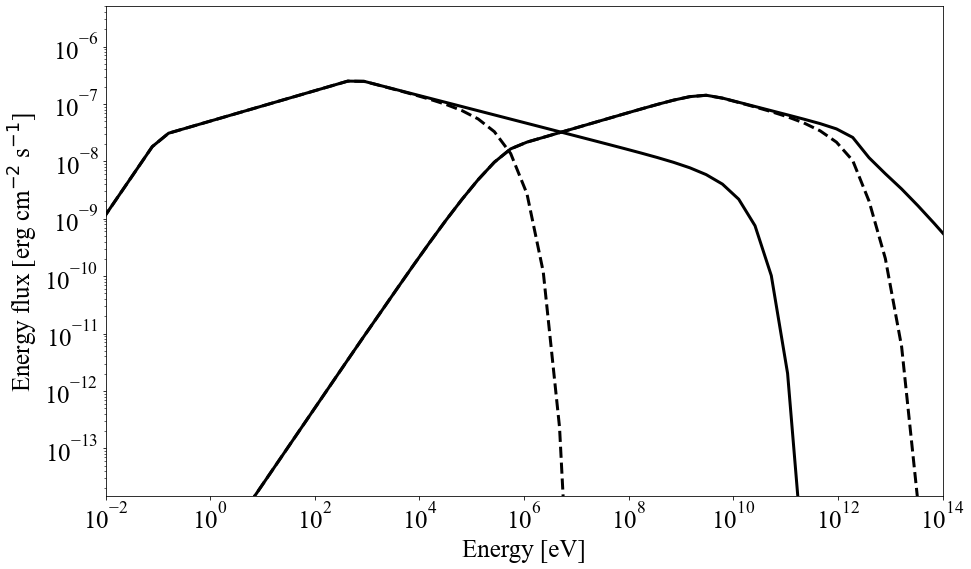}
    \label{fig:sed_gammamax}}\\
    \subfloat[]{\includegraphics[width=\columnwidth]{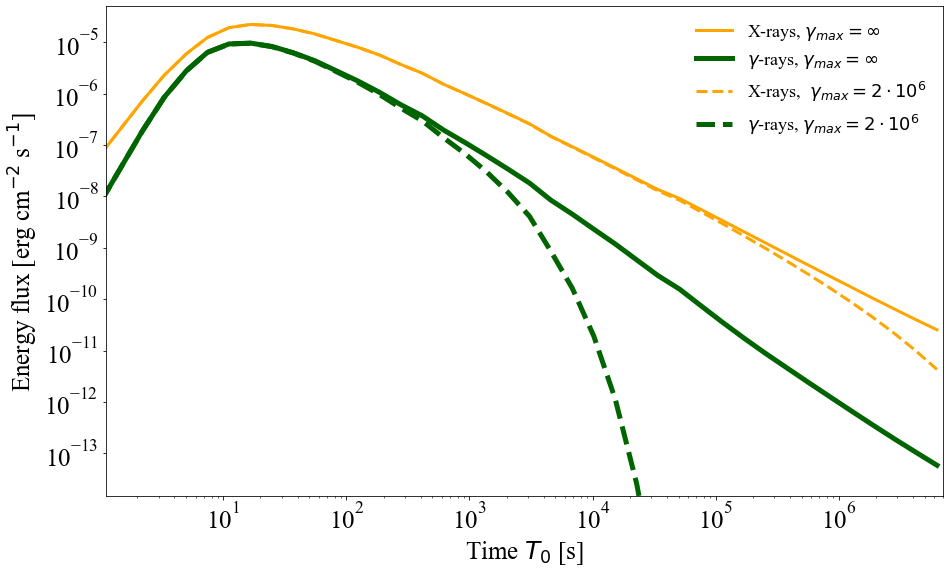}\label{fig:lightcurve_gammamax}}\\
    \caption{Calculated spectral (a) and flux evolution (b) of the SSC emission from a relativistic fireball for two different maximum energies of the electron distribution $\gamma_{\text{max}}=2 \cdot 10^6$ (dashed) and $\gamma_{\text{max}}=\infty$ (continuous line). In (b) we show  the calculated light curves in X-rays (thin blue lines) and VHE gamma~rays (thick red lines).  }
    \label{fig:model_gammamax_comparison}
\end{figure}

\subsection{Diagnostics at TeV gamma-ray energies}
\label{sec:diagnostics}
\noindent
TeV gamma-ray observations represent a great opportunity to investigate the physics underlying GRBs. In particular, they provide a powerful diagnostic tool for probing the maximum energy cutoff in the particle distribution $\gamma_{\text{max}}$ responsible for producing the SSC emission.\\

\noindent
\begin{center}
\textit{Probing the maximum particle energy $\gamma_{\text{max}}$}\\
\end{center}An illustrative example on the importance of the $\gamma_{\text{max}}$ parameter is presented in \Cref{fig:model_gammamax_comparison}, which compares two sets of spectra and light curves derived from the relativistic fireball model, highlighting the effects of a finite $\gamma_{\text{max}}$ in the electron distribution. Dashed lines represent the model with $\gamma_{\text{max}} = 2 \cdot 10^6$, while continuous lines describe a model with $\gamma_{\text{max}} \sim \infty$, typical of standard GRB models \citep[e.g.,][]{sari_1998}.

The upper panels show the resulting broad-range spectra resulting from such models. As expected, the model with lower $\gamma_{\text{max}}$ produces spectra with a minor extension of the synchrotron and IC emission at high energies. This behavior is also evident in the lower panels, which show the computed lightcurves. The earlier spectral cutoff associated with the lower $\gamma_{\text{max}}$ - with respect to the cutoff at the synchrotron burn-off limit in the case $\gamma_{\text{max}} \sim \infty$ - causes a corresponding steepening in the flux evolution for both the X-ray and the TeV gamma-ray energy bands. Notably, the TeV gamma-ray light curve shows this steepening earlier than the corresponding X-ray lightcurve. 

For this reason, the relevance of a maximum $\gamma_{\text{max}}$ in the accelerated particle distribution represents an important feature of GRB modeling. Appreciating this point
was challenging in the past due to the very late appearance in the X-ray lightcurve, which may also be contributed by other effects at late times. Conversely, the new opportunities offered by the TeV gamma-ray observations - even when performed at one or two days from the trigger event - may test and lead to a determination of such a maximum energy of the particle distribution $\gamma_{\text{max}}$.
This is a key feature emerging now with TeV gamma-ray observations of GRBs, such as GRB~221009A \citep[which we discussed in ][]{foffano2024_GRB221009A} and probably also in the case - less constrained - of GRB~180720B.

It is also worth mentioning the importance of simultaneous MeV and GeV observations in improving our understanding on the particle distribution underlying long GRBs. Specifically, this energy band can show the  emergence of a spectral break at MeV energies with respect to the upcoming IC flux at GeV energies, defined by a finite maximum particle energy $\gamma_{\text{max}}$.
This feature is clearly illustrated in \Cref{fig:sed_gammamax}.
Simultaneous observations of this energy band will pinpoint this physical feature of GRBs, although this effect may become evident only in favorable GRB events and typically several hundred seconds after the trigger, once the initially high Lorentz factor causes the synchrotron cutoff to shift down into this energy range. \\

\subsection{Expectations for early and late TeV observations of gamma-ray bursts}
\noindent
Early TeV gamma-ray observations at 10-100~s after the trigger are crucial for probing the physics during the transition from the prompt phase to the early afterglow phase and identifying the earliest gamma-ray signal of GRBs. However, such early observations are complicated by the intrinsic reaction times of Cherenkov observatories, which are currently standing just below a minute, but in the future they may be lowered to a few tens of seconds after the GRB trigger announcement. Of course, gamma-ray observatories with significantly better duty cycles, such as LHAASO, are not limited by this problem, although their observing performance is generally worse.

Observations at one or two days after the trigger time can provide invaluable physical insights on long GRBs.  
Longer exposure times available for these observations imply more favorable lower sensitivities of the TeV Cherenkov instruments.
As shown in \Cref{fig:lightcurve_comparisons_new_nodata}, these observations directly contribute to our knowledge of the maximum particle energy $E_{\text{max}}$. 

Indeed, depending on $\gamma_{\text{max}}$, there are two different behaviors of the late TeV afterglow lightcurves. If $\gamma_{\text{max}}$ is relatively large, $\gamma_{\text{max}} \sim 10^8 - 10^9$, the late-time TeV light curves can be represented as temporal power law functions with small, if any, curvatures. \Cref{fig:lightcurve_comparisons_new_nodata} shows that this behavior can be identified for GRB~190114C, GRB~190829A, and GRB~201216C.

On the contrary, significant curvature in the late TeV light curve is expected for relatively small values of $\gamma_{\text{max}} \sim 10^6-10^7$, as in the cases of GRB~180720B and GRB~221009A.

Obviously, an ideal combination of early (10-100 s), intermediate (hours) and late (1-2 days) TeV observations can determine the exact shape of the TeV light curve and therefore, as we show in this paper, contribute to the determination of $\gamma_{\text{max}}$. As we show in \Cref{fig:lightcurve_comparisons_onlydata_eblabs}, this prospect for TeV Cherenkov telescopes is achievable today with the existing TeV instruments (MAGIC, H.E.S.S., VERITAS, HAWC, LHAASO), and it will be even more strengthened by future more sensitive TeV arrays (CTAO).\\

%%%%%%%%%%%%%%%%%%%%%%%%%%%%%%%%%%%%%%%%%%%%%%%%%%%%%%%
\section{Conclusions}
\noindent
In this work, we investigated the physical properties of five GRBs with afterglows detected at TeV energies. 
We modeled the available multi-frequency datasets employing an optimized relativistic fireball model, which offers an accurate description of the temporal and spectral properties of the five GRB afterglows. 
Our main goal was the comparison between the TeV gamma-ray datasets with the best-fitting light curves of the relativistic fireball model, especially at intermediate and late times.
In doing so, we globally modeled the medium to late-time afterglows, and considered optical, X-ray, GeV, and TeV light curves and spectra.

Our results show that the TeV afterglow light curves of GRB~221009A and possibly of GRB~180720B are affected by a steepening between $10^4$ and $10^6$~s. These decays can be caused by different effects, either jet breaks (in the simple case or also in the structured jet scenario) or intervening spectral effects.
In this paper, we advocate the latter mechanism that ultimately depends on the physical properties of the afterglows. In particular, we find that values of $\gamma_{\text{max}}$ in the range $10^6 - 10^7$ (instead of the typical values reported in the literature of $\gamma_{\text{max}} \sim \infty$) are necessary to successfully model light curves and spectra of GRB~221009A and GRB~180720B. This is one of the main conclusions of our paper.

Our results indicate that the TeV gamma-ray afterglows can be determined with an accuracy that can discriminate among different emission modes.
While early observations can investigate the transition from
the prompt phase to the earliest afterglow phase, later observations - extending up to a few days after trigger - offer the opportunity to explore important features of the physics underlying the extreme particle acceleration in long GRBs.  In particular, the TeV light curve steepening - due to the non-trivial values of $\gamma_{\text{max}}$ - can be detectable with gamma-ray observations even after one or two days after trigger. 
These late-time observations by TeV instruments can take advantage of longer exposure times compared to early-time observations, leading to improved sensitivity. These late observations may compensate for potential non-optimal atmospheric conditions at the trigger time of the events (an unavoidable limitation for ground-based observations).

Our results emphasize the relevance of late-time TeV observations, especially in light of the forthcoming generation of Cherenkov telescopes, which will benefit of much improved sensitivity in the TeV gamma-ray band.  Furthermore, obtaining spectra in the MeV-GeV energy range during the early and late afterglow phases is absolutely crucial to the theoretical modeling of the fascinating GRBs. Hopefully, a new generation of space gamma-ray instruments will be approved in the next years. Our paper shows that even without simultaneous MeV-GeV information, early and late afterglow TeV observations can validly compensate for the absence.  \\

\begin{acknowledgments}
\noindent
{Acknowledgments}\\
{We acknowledge the useful comments by the anonymous referees whose indications contributed to improve our paper.}
AGILE is a mission of the Italian Space Agency (ASI), with scientific and programmatic participation of INAF (Istituto Nazionale di Astrofisica) and INFN (Istituto Nazionale di Fisica Nucleare). 
This work was partially supported by the grant Addendum n.7 - Accordo ASI-INAF n. I/028/12/0 for the AGILE project.
\end{acknowledgments}

%%%%%%%%%%%%%%%%%%%%%%%%%%%%   BIBLIO    %%%%%%%%%%%%%%%%%%%%%%%%%%%%%%%%%%%%%%%%%%%%%%%%%%
\newpage
\bibliography{biblio}{}
\bibliographystyle{aasjournal}

\appendix

\section{Comparison of the evolution of the critical energy}
\label{sec:appendix_nu_c}
\begin{figure}[!ht]
    \centering
    \includegraphics[width=0.6\textwidth]{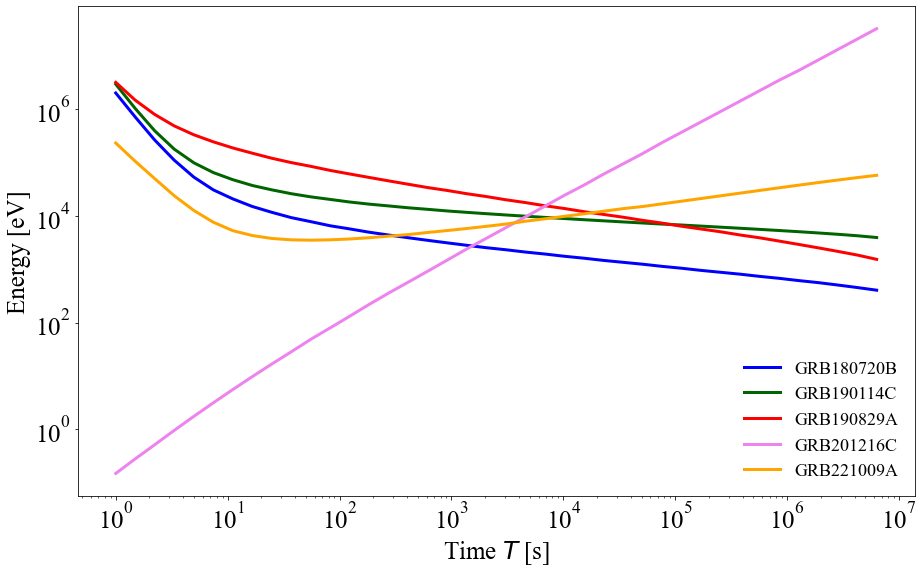}
    \caption{Time evolution of the energy corresponding to the cooling frequency $\nu_c$ of the model of the afterglow of each GRBs in the sample of \Cref{tab:list_grbs}. We remind that two non-standard cases are presented: GRB~201216C has been modeled with a wind-like density profile for the external medium and that GRB~221009A has been modeled with time-dependent microphysical efficiencies $\epsilon_e$ and $\epsilon_B$. In this plot, observational data are reported from the trigger time $T \equiv T_0$ for all GRBs, except for GRB~221009A for which $T \equiv T' = T_0 + 226$~s.}
    \label{fig:frequencies}
\end{figure}

\noindent
In \Cref{fig:frequencies}, we present the computed time evolution of the critical energy $E_c = h \nu_c$, where the latter is the critical frequency of the model corresponding to the critical energy $\gamma_c$, as discussed in Section~\ref{sec:modeling}. The typical evolution described by \citet{sari_1998} is reported, along with GRB~201216C which was modeled assuming a wind-like density profile of the external medium. GRB~221009A is reporting a peculiar evolution of the critical energies due to the specific  modeling with time-evolving microphysical efficiencies which include further time dependencies.

\end{document}